\documentclass[lettersize,journal]{IEEEtran}

\usepackage{bm}
\usepackage{stackrel}

\usepackage{amsmath,amsfonts}
\usepackage{algorithmic}
\usepackage{algorithm}
\usepackage{array}
\usepackage{textcomp}
\usepackage{stfloats}
\usepackage{url}
\usepackage{verbatim}
\usepackage{graphicx}
\usepackage{caption}
\usepackage{subcaption}
\usepackage{cite}

\usepackage{scalerel}
\usepackage{tikz}
\usetikzlibrary{svg.path}

\definecolor{orcidlogocol}{HTML}{A6CE39}
\tikzset{
  orcidlogo/.pic={
    \fill[orcidlogocol] svg{M256,128c0,70.7-57.3,128-128,128C57.3,256,0,198.7,0,128C0,57.3,57.3,0,128,0C198.7,0,256,57.3,256,128z};
    \fill[white] svg{M86.3,186.2H70.9V79.1h15.4v48.4V186.2z}
                 svg{M108.9,79.1h41.6c39.6,0,57,28.3,57,53.6c0,27.5-21.5,53.6-56.8,53.6h-41.8V79.1z M124.3,172.4h24.5c34.9,0,42.9-26.5,42.9-39.7c0-21.5-13.7-39.7-43.7-39.7h-23.7V172.4z}
                 svg{M88.7,56.8c0,5.5-4.5,10.1-10.1,10.1c-5.6,0-10.1-4.6-10.1-10.1c0-5.6,4.5-10.1,10.1-10.1C84.2,46.7,88.7,51.3,88.7,56.8z};
  }
}

\newcommand\orcidicon[1]{\href{https://orcid.org/#1}{\mbox{\scalerel*{
\begin{tikzpicture}[yscale=-1,transform shape]
\pic{orcidlogo};
\end{tikzpicture}
}{|}}}}

\usepackage{hyperref} 

\begin{document}

\title{Approximate Wireless Communication for Lossy Gradient Updates in IoT Federated Learning}
\author{Xiang Ma,$^{\textsuperscript{\orcidicon{0000-0003-0401-7101}}}$\,\IEEEmembership{Student Member,~IEEE,} Haijian Sun,$^{\textsuperscript{\orcidicon{0000-0002-0680-147X}}}$\,\IEEEmembership{Member,~IEEE,} \\ Rose Qingyang Hu,$^{\textsuperscript{\orcidicon{0000-0002-1571-3631}}}$\,\IEEEmembership{Fellow,~IEEE,} and Yi Qian,$^{\textsuperscript{\orcidicon{0000-0001-5671-916X}}}$\,\IEEEmembership{Fellow,~IEEE}

\thanks{Manuscript received April xx, 2024; revised June xx, 2024. This article was partly presented at the ACM Workshop on Wireless Security and Machine Learning (WiseML 2023).}
\thanks{Xiang Ma and Rose Qingyang Hu are with the Department of Electrical and Computer Engineering, Utah State University, Logan, UT 84322 USA (e-mail:
xiang.ma@ieee.org, rose.hu@usu.edu).}
\thanks{Haijian Sun is with the School of Electrical and Computer Engineering,
University of Georgia, Athens, GA 30602 USA (e-mail: hsun@uga.edu).}
\thanks{Yi Qian is with the Department of Electrical and Computer Engineering, University of Nebraska–Lincoln, Omaha, NE 68182 USA (e-mail:
yi.qian@unl.edu).}
}



\maketitle

\begin{abstract}
Federated learning (FL) has emerged as a distributed machine learning (ML) technique that can protect local data privacy for participating clients and improve system efficiency. Instead of sharing raw data, FL exchanges intermediate learning parameters, such as gradients, among clients. This article presents an efficient wireless communication approach tailored for FL parameter transmission, especially for Internet of Things (IoT) devices, to facilitate model aggregation. Our study considers practical wireless channels that can lead to random bit errors, which can substantially affect FL performance. Motivated by empirical gradient value distribution, we introduce a novel received bit masking method that confines received gradient values within prescribed limits. Moreover, given the intrinsic error resilience of ML gradients, our approach enables the delivery of approximate gradient values with errors without resorting to extensive error correction coding or retransmission. This strategy reduces computational overhead at both the transmitter and the receiver and minimizes communication latency. Consequently, our scheme is particularly well-suited for resource-constrained IoT devices. Additionally, we explore the inherent protection of the most significant bits (MSBs) through gray coding in high-order modulation. Our simulations demonstrate that our proposed scheme can effectively mitigate random bit errors in FL performance, achieving similar learning objectives, but with the $50\%$ air time required by existing methods involving error correction and retransmission. 


\end{abstract}


\section{Introduction}
Federated learning (FL) \cite{fl} is an emerging learning paradigm that allows clients to perform machine learning (ML) tasks locally while benefiting from the collective learning capabilities of other clients through the exchange of model parameters. Specifically, client data remains locally and only the learned model is shared, ensuring robust privacy protection for all participants. The FL system comprises a central parameter server (PS) and multiple clients. The server aggregates the local models uploaded by the clients and sends the global model back. This iterative learning process continues round by round, where the results of the previous round establish the foundation for the subsequent ones. Within each round, the server disseminates the current global model to clients, who then perform local computations using their environment-specific data and the latest received global model. Subsequently, the new local models are uploaded to the server for aggregation. This cyclical procedure persists until the global model converges.
\IEEEpubidadjcol

For edge devices such as unmanned aerial vehicles (UAVs) and IoT devices such as smart watches, wireless networks are commonly used to connect them to the server \cite{fl_uav}. However, the nature of random wireless channels can introduce errors during data transmission. Numerous methods have been proposed to address this issue. For example, in the physical layer, high-power transmission can enhance signal quality and overcome noise effects; receiver equalization can mitigate the impact of fading channels, etc. However, this approach may cause communication overhead. In the upper layers, error detection and correction methods such as parity check, checksum check, and cyclic redundancy check (CRC), and forward error correction (FEC) \cite{fec} are commonly utilized. FEC encodes the message with redundant bits as an error correction code (ECC) to correct certain errors on the receiver side. Common ECCs for error correction include convolutional code, low-density parity check (LDPC) code, and turbo code. However, it introduces redundancy and requires additional computations on the transmitter and receiver for encoding and decoding the message \cite{fec_overhead}. Furthermore, communication overhead is introduced to transmit redundancy information. There exists a trade-off between error correction capability and computation/communication efficiency. When the channel conditions are poor, errors may exceed the ECC's correction capability. In such cases, packet retransmission is employed in the transport layer to ensure reliable transmission.  FEC and packet retransmission can intensely drain energy-constrained IoT devices \cite{battery_drain}.  A study by the authors in \cite{fl_error} focused on transmission bit errors in FL, specifically in a packet erasure channel. The server discards erroneous local models in this scenario and resorts to past local models for continuity. This will cause the most recent model updates to be lost. In \cite{lossy_transmission}, the authors proposed the FedLC framework, which applied the user datagram protocol (UDP) rather than the transmission control protocol (TCP) as the basis for transmitting the model update in a lossy communication channel. FEC and packet retransmission are also used as countermeasures to prevent packet losses.

Numerous schemes have been devised in FL to mitigate communication overhead. To allow large-step model updates to be uploaded at once, Federated Averaging (FedAvg) \cite{fl_sgd} groups multiple stochastic gradient descent (SGD) updates together. However, transmitting gradients can still lead to significant delays for large-scale distributed ML models with millions of parameters. In \cite{fl_convergence_time} and \cite{scheduling_policy_power}, a joint problem of resource allocation and user selection is proposed to minimize the convergence time in FL. Advanced transmission schemes such as non-orthogonal multiple access (NOMA) \cite{noma} offer promising solutions. NOMA allows multiple users in the FL system to transmit model parameters simultaneously, and the server decodes the superposition signal using successive interference cancelation (SIC), thus improving channel efficiency and reducing communication delay. Another practical approach to address this challenge is gradient compression, with extensive research demonstrating the efficacy of gradient sparsification and quantization while incurring minimal performance loss. For example, 1-bit SGD was applied in \cite{one_bit} to reduce the transmission size of the gradient in a distributed data-parallel system, achieving at least $4 \times$ reduction in the real training time from end to end. Furthermore, in \cite{sparse}, the authors found that the drop in $99\%$ of the gradients results in no or negligible loss of precision in multiple data sets. A joint learning and communication problem is formulated in \cite{fl_energy_efficient} to improve the energy efficiency for IoT devices. 

Approximate communication, an analogy of approximate computing techniques in parallel systems, reduces communication overhead between transmitters and receivers \cite{approx_computing}. This approach permits communications to be executed ``approximately", deliberately allowing minor errors to achieve efficiency gains. However, the system must possess intrinsic error resiliency, and the error should be acceptable in order to adopt approximate communication successfully. Within the FL system, error resiliency manifests itself in various aspects. First, the gradients in ML exhibit minimal variation, as demonstrated by empirical studies in \cite{terngrad, statistical, grad_comp}. These studies reveal that gradient values often fall within the $(-1, 1)$ or even narrower, such as $(-0.01, 0.01)$. \cite{terngrad} provides histograms of gradients for fully connected and convolutional layers in different training iterations, while \cite{statistical} presents probability density functions (PDFs) and cumulative distribution functions (CDFs) of gradients. The gradients are Gaussian or near-Gaussian distributed. These sources collectively establish that the ML gradients are distributed within a small range and that this range is empirically known. Furthermore, the error resilience of the FL system is demonstrated through gradient compression, which introduces errors in the form of gradient quantization. Individual gradient errors arise during quantization. However, while individual gradients are accurately represented, a fraction of all gradients are removed through gradient sparsification. Finally, the FL system's model aggregation occurs on average, enhancing error resilience by limiting errors to an acceptable level. With more clients participating in the learning process, the resilience to errors improves.

Approximate communication has been applied in various domains, such as network-on-chip (NoC) designs, contributing significantly to higher energy efficiency \cite{arxon, chen, approx_strategy}. For example, in \cite{arxon}, an approximate communication framework was designed for photonic NoCs to reduce the overhead of lasers and turning power while maintaining acceptable distortion levels. This approach achieved an impressive reduction of $56.4\%$ in power consumption compared to the best-known prior work. Furthermore, \cite{chen} introduced a quality control method that collaborates with data approximation mechanisms to reduce packet size, thus reducing network power consumption and latency. The quality control method identifies error-resistant variables and calculates error thresholds based on quality requirements. \cite{approx_strategy} proposed packet production and reduction in error checking for NoC as approximate communication solutions. The proposed scheme achieves better performance in both energy consumption and latency. Approximate communication was also presented in \cite{media} to eliminate the need for additional network or spectrum resources for redundant information transmission in wireless media delivery. By prioritizing significant bits in wireless packets and placing them in the most significant bit positions during high-order modulation, the proposed method significantly improved video quality by 5 to 20 dB under wireless conditions. Moreover, in \cite{approx_distribute}, researchers presented a distributed approximate Newton-type method to improve communication efficiency in stochastic optimization and learning problems, accelerating the convergence speed of quadratic problems. This demonstrates the broad applicability and potential benefits of approximate communication techniques in various fields and scenarios.

Inspired by these works, we introduce an approximate wireless communication framework for FL. This framework encourages the lossy transmission of FL model gradients, offering the advantage of low latency, reduced overhead, and reduced computation. Unlike \cite{lossy_transmission}, the proposed approximate wireless communication operates in the physical layer, while FedLC uses FEC in the application layer and packet retransmission, UDP in the transport layer. They can work together to achieve better performance, but FEC and packet retransmission will cause extra costs.

In this research, we conducted a comprehensive theoretical analysis of ML gradients within widely used ML setups. Specifically, we analyze the gradients in fully connected neural networks and convolutional neural networks (CNN) under commonly employed structures like Sigmoid/ReLU activation and cross-entropy loss functions. This enables us to clarify the gradient vanishing and gradient exploding problems in deep neural networks (DNN). The gradient clipping constraints gradient to a working range and mitigates the gradient exploding problems. Using this insight, we set a threshold for erroneous gradients in conjunction with approximate transmission in practical wireless networks. Through extensive simulations, we substantiate the effectiveness of our proposed method in mitigating the impact of bit errors on FL performance. Our approach saves significant time, reducing transmission duration by half compared to traditional Error Correction and ReTransmission (ECRT) techniques. The main contributions of this study are summarized below.
\begin{itemize}
    \item[1)] This study offers a sketch mathematical analysis of gradient values in commonly employed ML settings. We demonstrate that back-propagation could result in gradient vanishing and gradient exploding problems.

    \item[2)] Using prior knowledge of gradients and the effectiveness of gradient clipping, received bits are thoughtfully masked to confine errors within a small range. This allows approximate transmission for error-resilient FL systems.  

    \item[3)] Approximate communication finds practical application in wireless gradient transmission, effectively reducing communication latency and computational overhead. Additionally, using gray coding with high-order modulation offers protection for the most significant bits (MSBs), resulting in notable improvements in learning performance. Gradient sparsification can further reduce communication costs and mitigate the side effects caused by approximate communication.

    \item[4)] Extensive simulations are conducted to validate the efficacy of the proposed approximate wireless communication for FL gradients. Learning performance is evaluated by considering factors such as the number of users, the size of the error, and the modulation index that can influence the results.
\end{itemize}

The rest of the paper is organized as follows. Section II provides an introduction to the system model, including both the FL system model and the wireless channel model. Section III presents the mathematical analysis of gradients in commonly used ML settings. Section IV describes the proposed approximate transmission method. Section V presents the simulation results, and finally, Section VI concludes the paper.

\section{System Model}
In this section, we begin by introducing the FL system, followed by the wireless channel model. Specifically, we consider an FL system comprising $M$ edge clients, each connected to the server wirelessly. The entire dataset $D$ is distributed among these $M$ clients, each client $m$ containing a subset of data denoted as $D_m$, i.e., $\sum_m |D_m| = |D|$, where $|D_m|$ and $ |D|$ represent the size of dataset $ D_m $ and $D$.

\subsection{FL System}
\begin{figure}[ht]
	\includegraphics[width=3.0in]{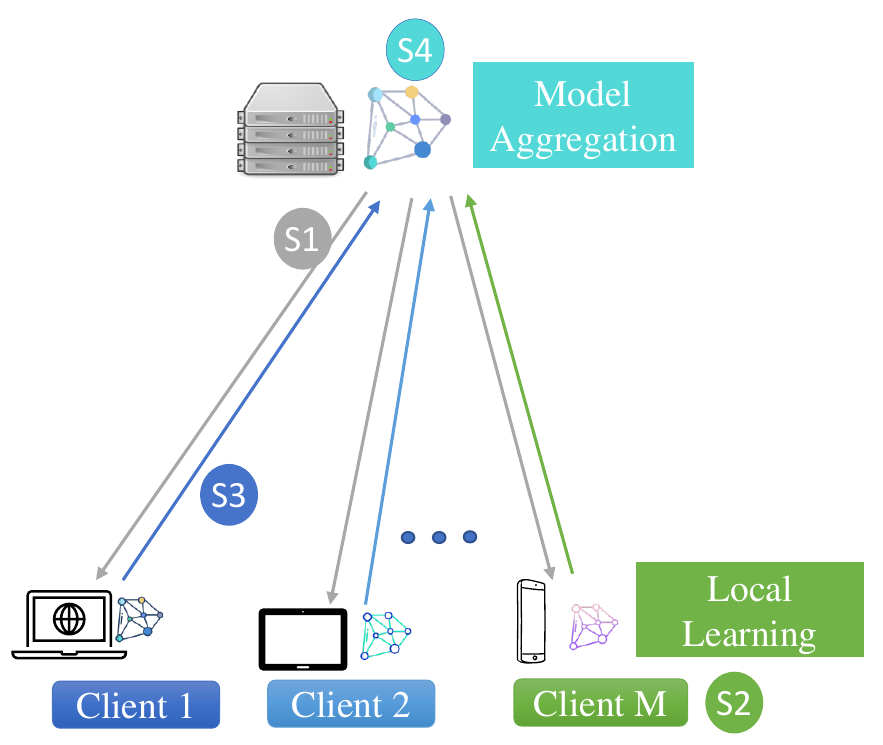}
	\centering
	\caption{FL System Model}
	\label{fig:system_model}
\end{figure}

FL operates as an iterative ML algorithm between the server and the clients, progressing round by round. Each round entails four fundamental steps. 
The system model is illustrated in Fig. \ref{fig:system_model}. Wireless communication is used for the downlink in step (S1) and the uplink in step (S3).

The objective function of FL can be defined as:
\begin{equation}
\min_{\bm{w}\in R^d} f(\bm{w}) \quad  \text{where} \quad f(\bm{w}) \stackrel{\text{def}}{=}\frac{1}{|D|}\sum_{i=1}^{|D|} f_i(\bm{w}), \label{eq:objective}
\end{equation} The function $f_i(\bm{w}) = C(\bm{x_i}, \bm{y_i};\bm{w})$ serves as the cost or loss function, quantifying the inference error between the data sample $(\bm{x_i}, \bm{y_i})$ and the inference produced by model parameters $\bm{w}$. The commonly employed loss function for classification problems in ML is the cross-entropy function, particularly in neural network models. In multiclass classification, the label $\bm{y_i}$ is often one-hot encoded, ensuring that each label carries equal weight. Thus, $\bm{y_i}=[0,\dots,0,1,0,\dots,0]^T$, with the value $1$ located at the $i$-th position.

When the data are distributed among $M$ clients, the objective function (\ref{eq:objective}) can be rewritten as follows:
\begin{equation}
    f(\bm{w}) = \sum_{m=1}^M \frac{|D_m|}{|D|} F_m(\bm{w}), 
\end{equation}
where $F_m(\bm{w}) = \frac{1}{|D_m|}\sum_{i\in D_m} f_i(\bm{w})$. Distributing data and computation across multiple clients can convert a traditional centralized ML problem into a distributed FL problem.

Given that the cost function for neural networks is typically non-convex, directly solving for the global minimum poses challenges. Consequently, the gradient descent method is commonly used in ML to identify local minimum points. Stochastic gradient descent (SGD), a variant of the gradient descent method, is beneficial in avoiding local minimums by randomly selecting data samples. As a result, gradients play a central and crucial role in the learning process. The gradient is defined as
\begin{equation}
    g = \nabla_{\bm{w}} C(\bm{x_i}, \bm{y_i};\bm{w}).
\end{equation}
The gradient represents the derivative of the loss function $C(\cdot)$ concerning the model parameters $\bm{w}$. In each round, the local gradient at each client can be expressed as follows:
\begin{equation} \label{eq:local_comp}
    g_t^m = \nabla F_m(w_t).
\end{equation}
$g_t^m$ is the local gradient of the client $m$ in the round $t$. The global gradient after aggregation in round $t$ is 
\begin{equation}
    g_t = \sum_{m=1}^M \frac{|D_m|}{|D|} g_t^m.
\end{equation}
$\frac{|D_m|}{|D|}$ denotes the aggregation coefficient, ensuring that $\sum_m \frac{|D_m|}{|D|} = 1$. The server maintains the model weights of the last round, denoted $w_t$, and updates the global model using the following procedure:
\begin{equation} \label{eq:sgd_update}
    w_{t+1} = w_t - \eta g_t.
\end{equation}
Here, $\eta$ is the learning rate, typically within the $(0,1)$ range.

\subsection{Wireless Channel Model}
FL operates as an application layer algorithm that remains agnostic to the lower-layer gradient transmission specifications, irrespective of whether it involves fiber, cable, or radiowaves. Upper-layer applications directly utilize the data received from the lower layers, assuming reliable transmission. However, when UAVs or IoT devices engage in the FL learning process, transmission often occurs over wireless channels, characterized by dynamic and unpredictable behavior, making it inherently unreliable. To ensure reliable communication, additional measures such as FEC and packet retransmission are necessary to overcome the challenges of the wireless channel. In the context of FL, we consider the fading channel in the uplink, as illustrated in step (S3), which may lead to random bit errors. In contrast, for the downlink, we assume that the server can provide global gradients to clients with negligible errors, for example with higher transmit power that leads to a higher signal-to-noise ratio (SNR) \cite{fl_error}.

In the uplink, a time division scheme can be employed, where each user is assigned a specific time slot while sharing the same channel. The signal received on the server can be expressed as follows:
\begin{equation}
    r_t^m = \sqrt{p_t^m (d^m)^{-\alpha} } h_t^m g_t^m + n_t^m,
\end{equation}
where $r_t^m$ denotes the signal received at the server from client $m$ during round $t$. Transmission power is represented by $p_t^m$, and small-scale fading by $h_t^m$, assumed to follow a complex standard Gaussian distribution, that is, $h_t^m \sim \mathcal{CN} (0, 1)$. Large-scale fading is explained by path loss, the distance between the server and the client $m$ denoted as $d^m$, and the path loss exponent represented by $\alpha$. Additive noise is assumed to follow a Gaussian distribution, $n^t \sim \mathcal{CN} (0, n_\sigma^2)$, where $n_\sigma^2$ represents the variance. Additionally, we assume that the server has channel information, that is, $c_t^m=\sqrt{p_t^m (d^m)^{-\alpha} } h_t^m$, with only the noise remaining unknown. 

The transmission process can be described below. First, the gradients are converted from decimal to binary format. Subsequently, the bits are mapped to symbols using a quadrature amplitude modulation (QAM) scheme. The symbols are then transmitted through the wireless fading channel. At the receiver side, the signal is decoded and demodulated to the closest point in the QAM constellation. The demodulation process can be represented as follows:
\begin{equation}
    \hat{g}_t^m = \arg_{\Bar{g}_t^m \in \mathcal{G}} \min || r_t^m - \sqrt{p_t^m (d^m)^{-\alpha} } h_t^m \Bar{g}_t^m||^2,
\end{equation}
where $\mathcal{G}$ refers to the set of symbol points on the constellation diagram. $\Bar{g}_t^m$ represents a potential symbol point within $\mathcal{G}$, while $\hat{g}_t^m$ denotes the optimal symbol point.

The main notation used in the paper is summarized in Table I. 
\begin{table}[h]
	\newcommand{\tabincell}[2]{\begin{tabular}{@{}#1@{}}#2\end{tabular}}
	\centering
	\caption{Summary of Notations}
	\begin{tabular}{c|p{65mm}}
		\hline
		\textbf{Notation} & \textbf{Definition}\\
		\hline
		$M$; $m$ & The total number of clients connected to the server; the client index\\
		\hline
		$L$; $l$ & The last layer of the neural network; the neural network layer index \\
		\hline
		$\bm{x}$; $\bm{y}$; $\hat{y}$ & Features of a data point sample; corresponding true label of the data point; the predicted label\\
            \hline
            $w$; $g$; $b$ & Model weight; model gradient; neural network bias \\
            \hline
            $\eta$; $\delta$ & Learning rate; intermediate quantity as ``error'' \\ 
		\hline
            $C$; $\sigma(\cdot)$ & Loss function; activation function \\
            \hline
            $z$; $a$ & Intermediate output of neuron; neuron output \\
            \hline
            $N_l$; $N_g$; $N$ & The number of neurons in the $l$-th layer; the number of gradients in neural network; the number of data samples \\
            \hline
            $i$; $j$; $k$; $p$; $q$ & Index \\
            \hline
            $s$; $t$; $u$, $v$ & Index \\
            \hline
            $t$ & Time (round) index \\
            \hline
            $h$; $\alpha$; $n$; $r$ & Channel factor; path loss exponent; noise; received signal \\
            \hline
            $p_t^m$; $d^m$ & The transmission power of the client $m$ at time $t$; Distance between the client $m$ and the server \\
            \hline
    
	\end{tabular}
	\label{Tab:notation}
\end{table}

\section{Gradients Analysis with Back-propagation}
ML model training aims to find the optimal point to minimize loss functions. Since neural networks are usually non-convex, SGD and back-propagation are typical methods for model training. However, gradient vanishing or exploding problems can occur in deep neural networks, preventing efficient model learning. The transmission of gradients with errors can cause the gradient to change in random directions. The received gradients can be substantial, making the model unstable. Studying the gradient behavior and the existing method to prevent gradient vanishing/exploding helps to design the method to perform approximate communication.

\subsection{Gradient in Fully Connected Neural Networks}
\begin{figure}[ht]
	\includegraphics[width=3in]{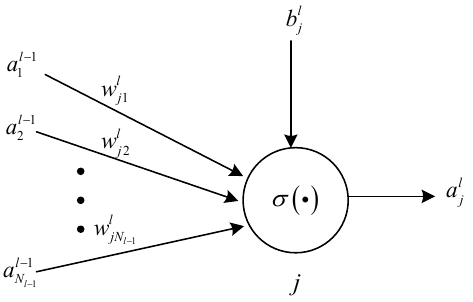}
	\centering
	\caption{Artificial Neuron Diagram}
	\label{fig:neuron}
\end{figure}

A neuron in neural networks serves as the fundamental unit for data processing, taking inspiration from neurons in human brains. It accepts input from the preceding layers, processes them, and transmits them to the next layer \cite{introduction_neuron}. The diagram depicting neurons is illustrated in Fig. \ref{fig:neuron}, providing insight into the input and output of the $j$-th neuron in the $l$ -th layer.

In the ML field, especially within neural networks, SGD and backpropagation are widely used optimization techniques for locating global minimum points. In a fully connected neural network, the feedforward equation at each neuron can be expressed as follows: 
\begin{equation} \label{eq:neuron_op}
\begin{aligned}
& z_j^l = b_j^l + \sum_k^{N_{l-1}} w_{jk}^l a_{k}^{l-1}, \\
& a_j^l  = \sigma(z_j^l).
\end{aligned}    
\end{equation}
Here, $b$ represents the bias, $w$ denotes the weights, $z$ stands for the intermediate output, and $a$ means the neuron output after passing through the activation function $\sigma(\cdot)$. The indices $j$ and $l$ denote that this neuron is the $j$-th neuron located in the $l$-th layer of the neural network. Additionally, $k$ is the index of the input from the $(l-1)$-th layer, and $N_{l-1}$ is the number of neurons in the $(l-1)$-th layer. In particular, the neuron output from the previous layers serves as the input to the current layer, and this process can continue back to the first layer, as illustrated in Fig. \ref{fig:neuron}. And equation (\ref{eq:neuron_op}) becomes
\begin{equation}
\begin{aligned}
& z_j^1 = b_j^1 + \sum_k^{N_{0}} w_{jk}^1 x_{k}^0, \\
& a_j^1  = \sigma(z_j^1).
\end{aligned}  
\end{equation}
in the first layer. $x_{k}^0$ is the $k$-th input, and the superscript $0$ is used to maintain writing consistency. The corresponding four fundamental equations in backpropagation for a fully connected neural network are \cite{nielsen}
\begin{subequations}
\begin{alignat}{2}
& \delta_j^L = \frac{\partial C}{\partial z_j^L} = \frac{\partial C}{\partial a_j^L} \frac{\partial a_j^L}{\partial z_j^L} = \frac{\partial C}{\partial a_j^L}  \sigma'(z_j^L) \label{eq:bp1},\\
& \delta_j^l = \frac{\partial C}{\partial z_j^l} = \sum_k^{N_{l+1}} \frac{\partial C}{\partial z_k^{l+1}} \frac{\partial z_k^{l+1}} {\partial a_j^l} \frac{\partial a_j^l}{\partial z_j^l} = \sum_k^{N_{l+1}} \delta_k^{l+1} w_{kj}^{l+1}\sigma'(z_j^l) \label{eq:bp2},\\
& \frac{\partial C}{\partial b_j^l} = \frac{\partial C}{\partial z_j^l} \frac{\partial z_j^l}{\partial b_j^l} = \delta_j^l \label{eq:bp3},\\
& \frac{\partial C}{\partial w_{jk}^l}  = \frac{\partial C}{\partial z_j^l} \frac{\partial z_j^l}{\partial w_{jk}^l} = \delta_j^l a_k^{l-1} \label{eq:bp4}.
\end{alignat}
\end{subequations}
Here, $L$ designates the final layer in the neural network, and $\delta_j^l$ signifies the ``error" in the $l$-th layer in the neuron $j$. $\delta_j^l$ serves as an intermediate quantity introduced essential for calculating the partial derivatives in equations (\ref{eq:bp3}) and (\ref{eq:bp4}).

The weight update can be rewritten as 
\begin{equation}
    w_{jk}^l = w_{jk}^l - \eta \delta_j^l a_k^{l-1}.
\end{equation}
Similarly, the bias update can be written as 
\begin{equation}
    b_j^l = b_j^l - \eta \delta_j^l.
\end{equation}

The calculation of the gradient $g^l = \nabla_{\bm{w^l}} C(\bm{x_i}, \bm{y_i};\bm{w^l})= [\frac{\partial C}{\partial w_{j_1 k_1}^l}, \frac{\partial C}{\partial w_{j_1 k_2}^l}, \dots, \frac{\partial C}{\partial w_{N_l N_{l-1}}^l}]$  in equation (\ref{eq:bp4}) involves two terms, $\delta_j^l$ and $a_k^{l-1}$. These two terms are discussed separately. Firstly, $a_k^{l-1}$ denotes the output of the activation function of the neuron $k$ in the $(l-1)$-th layer. Its range depends on the specific activation function used. For example, the Sigmoid function ensures that $a_k^{l-1}$ falls within the range $(0,1)$ regardless of input $z_k^{l-1}$. While for the ReLU activation function, the output of the activation function $a_k^{l-1}$ depends on the input $z_k^{l-1}$. More detailed analysis and mathematical expressions on activation functions can be found in \cite{activation}.

The calculation of the other term $\delta_j^l$ is described in equation (\ref{eq:bp2}), which involves a sum of products between the errors of the next layer $\delta_k^{l+1}$, the weights of the node $j$ in the $l$ -th layer to the node $k$ in the subsequent layer $w_{kj}^{l+1}$, and the derivative of the activation function $\sigma'(z_j^l)$. And the summation counts through all neurons in the $(l+1)$-th layer. The following analysis will go through the third term $\sigma'(z_j^l)$ first, then $w_{kj}^{l+1}$, finally $\delta_k^{l+1}$. 

Firstly, the derivative of the activation function $\sigma'(z_j^l)$ depends on the specific activation function used, with values ranging from $(0, 0.25)$ for the Sigmoid function and $\{0, 1\}$ for ReLU. Next, the weight $w_{kj}^{l+1}$ is based on the initialization of the model, the learning rate $\eta$, and the previous round gradients based on Equation (\ref{eq:sgd_update}). Common weight initialization methods generate random weight values within the range of $(-1,1)$ \cite{goodfellow}, drawn from Gaussian or uniform distributions \cite{initialization} in a heuristic manner. Furthermore, more recent initialization methods improve random initialization, such as in \cite{Xavier}, where the weight is initialized following a uniform distribution $w \sim U[\frac{1}{\sqrt{N_p}}, \frac{1}{\sqrt{N_p}}]$, $N_p$ is the number of neurons in the previous layer. The weight is in the $[\frac{1}{\sqrt{N_p}}, \frac{1}{\sqrt{N_p}}]$ range. And in \cite{he}, the weight is initialized in the Gaussian distribution, $w \sim \mathcal{N}(0, \sqrt{(2/N_p)})$, where $99.7\%$ of the weights are within $[3*\sqrt{(2/N_p)}, 3*\sqrt{(2/N_p)}]$. Lastly, the error $\delta_k^{l+1}$ for the third term can be expressed in the same way as in equation (\ref{eq:bp2}) with elements in the $(l+2)$-th layer as follows: 

\begin{equation}
\begin{aligned}
\delta_k^{l+1} = \frac{\partial C}{\partial z_k^{l+1}} & = \sum_i^{N_{l+2}} \frac{\partial C}{\partial z_i^{l+2}} \frac{\partial z_i^{l+2}} {\partial a_k^{l+1}} \frac{\partial a_k^{l+1}}{\partial z_k^{l+1}} \\
& = \sum_i^{N_{l+2}} \delta_i^{l+2} w_{ik}^{l+2}\sigma'(z_k^{l+1}). \label{eq:bp2_next}
\end{aligned}
\end{equation}
This process continues till the final layer.

In classification problems, the softmax function is commonly used as the activation function in the final layer to normalize the output class probabilities. It provides an effective combination when used in conjunction with the cross-entropy loss function. The cross-entropy loss function can be expressed as follows:
\begin{equation}
    C = -\sum_i  y_i \log(\hat{y}_i),
\end{equation}
where $y_i$ is the input truth label, $\hat{y}_i$ is the softmax probability for the $i$-th class, i.e.,
\begin{equation}
    \hat{y}_i = \sigma_s(z_i) = \frac{e^{z_i}}{\sum_k e^{z_k}}. 
\end{equation}
The derivative is 
\begin{equation}
\frac{\partial \hat{y}_i}{\partial z_j} = \left \{
  \begin{aligned}
    & \hat{y}_i (1-\hat{y}_j), && \text{if}\ i = j; \\
    & -\hat{y}_j \cdot \hat{y}_i, && \text{if} \ i \neq j. 
  \end{aligned} \right.
\end{equation}
Equation (\ref{eq:bp1}) can be written as 
\begin{equation}
\begin{aligned}
\delta_j^L = \frac{\partial C}{\partial \hat{y}_i^L} \frac{\partial \hat{y}_i^L}{\partial z_j^L} & = -\sum_i y_i \frac{\partial \log(\hat{y}_i)}{\partial \hat{y}_i} \frac{\partial \hat{y}_i}{\partial z_j}, \\
        & = -\sum_i y_i \frac{1}{\hat{y}_i} \frac{\partial \hat{y}_i}{\partial z_j}, \\
        & = -y_j(1-\hat{y}_j) - \sum_{i \ne j} y_i \frac{1}{\hat{y}_i}(-\hat{y}_j \cdot \hat{y}_i), \\ 
        & = \hat{y}_j \cdot \sum_i y_i - y_j.
\end{aligned}
\end{equation}

Now, equation (\ref{eq:bp4}) can be written as:
\begin{equation} \label{eq:final_fc}
    \begin{aligned}
        \frac{\partial C}{\partial w_{jk}^l} & = \delta_j^l a_k^{l-1} \\
        & = [\sum_p^{N_{l+1}} \delta_p^{l+1} w_{pj}^{l+1}\sigma'(z_j^l)] a_k^{l-1} \\
        & = \{ \sum_p^{N_{l+1}} [\sum_q^{N_{l+2}} \delta_q^{l+2} w_{qj}^{l+2}\sigma'(z_j^{l+1})] w_{pj}^{l+1}\sigma'(z_j^l) \} a_k^{l-1} \\
        & = \cdots \\
        & = \sum_p^{N_{l+1}} \cdots [\sum_i^{N_L} \delta_i^L w_{ij}^L\sigma'(z_j^{L-1})] \times \cdots \times a_k^{l-1}
    \end{aligned}
\end{equation}
And for the first layer, it becomes 
\begin{equation} \label{eq:fc_gradient_series}
\frac{\partial C}{\partial w_{jk}^1} = \sum_p^{N_{l+1}} \cdots [\sum_i^{N_L} \delta_i^L w_{ij}^L\sigma'(z_j^{L-1})] \times \cdots \times x_k^0
\end{equation}

As shown in Equation (\ref{eq:final_fc}) the gradient calculation involves multiplications and summation. When $\sum_p^{N_{l+1}} \delta_p^{l+1} w_{pj}^{l+1}\sigma'(z_j^l)$  for each layer is greater than 1, the multiplication increases the gradient even more. This can result in gradient exploding problems. On the contrary, when the summation is close to 0, the multiplication makes the gradient even smaller. And gradient vanishing problems could occur.

\subsection{Gradient in Convolutional Neural Networks}
\begin{figure}[ht]
	\includegraphics[width=3.4in]{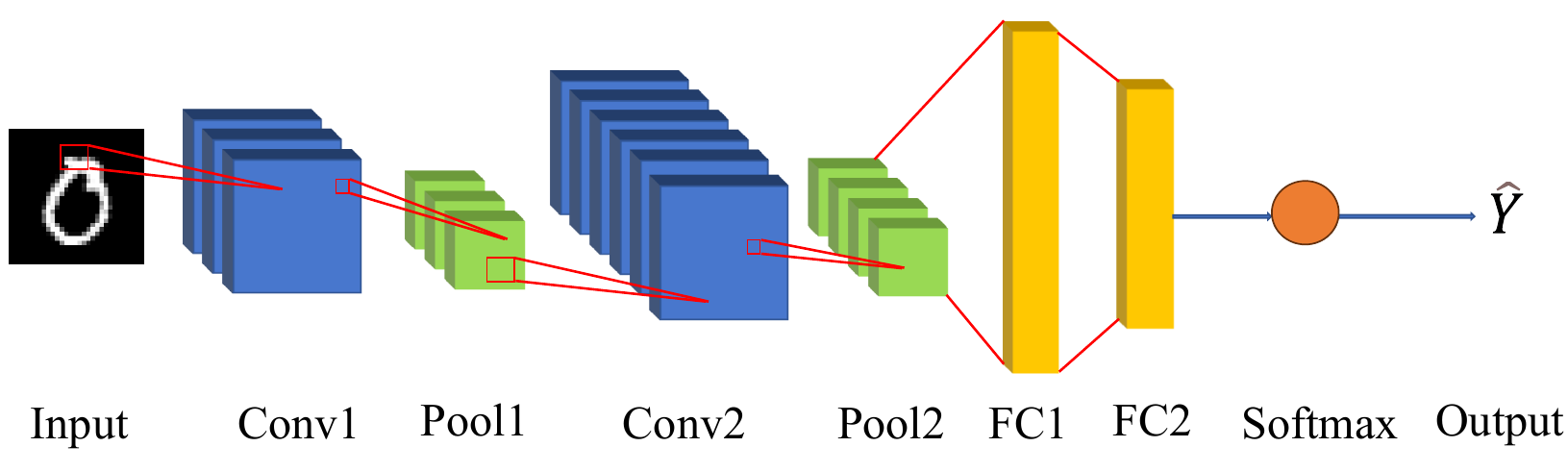}
	\centering
	\caption{Convolutional Neural Network Diagram}
	\label{fig:cnn_diagram}
\end{figure}

Modern image recognition tasks leverage CNNs as the primary technique. CNN is a particular variant of feedforward networks comprising three types of layers: convolutional, pooling, and fully connected. Typically, each convolutional layer is followed by a pooling layer for feature learning, and the network culminates with fully connected layers for classification \cite{convolutional}. Without loss of generality, the CNN network consists of two sets of typical CNN layers: two convolutional layers, two pooling layers, and two fully connected layers, which can be readily extended to a more general CNN structure. Moreover, we assume that the dimensions of the layers match, enabling practical training of the CNN. The diagram of this particular CNN is shown in Fig. \ref{fig:cnn_diagram}.

The feedforward process \cite{cnn_feedforward} can be written as:
\begin{subequations}
\begin{alignat}{2}
& z_{j,k}^1  = b_{j,k}^1 + \sum_p \sum_q w_{p,q}^1 x_{j+p, k+q}^0 \label{eq:cnn_conv1_z}, \\
& a_{j,k}^1  = \sigma(z_{j,k}^1) \label{eq:cnn_conv1_a}, \\
& a_{j,k}^2  = \max(a_{2j,2k}^1, a_{2j+1,2k}^1, a_{2j,2k+1}^1, a_{2j+1,2k+1}^1) \label{eq:cnn_pool1},\\
& z_{j,k}^3  = b_{j,k}^3 + \sum_p \sum_q w_{p,q}^3 a_{j+p, k+q}^2, \label{eq:cnn_conv2_z}  \\
& a_{j,k}^3  =  \sigma(z_{j,k}^3) \label{eq:cnn_conv2_a}, \\ 
& a_{j,k}^4  = \max(a_{2j,2k}^3, a_{2j+1,2k}^3, a_{2j,2k+1}^3, a_{2j+1,2k+1}^3) \label{eq:cnn_pool2},\\
& z_i^5      = b_i^5 + \sum_{j,k} w_{i;j,k}^5 a_{j,k}^4 \label{eq:cnn_fc1_z},\\
& a_i^5      = \sigma(z_i^5) \label{eq:cnn_fc1_a}, \\
& z_i^6      = b_i^6 + \sum_{k} w_{i,k}^6 a_{k}^5 \label{eq:cnn_fc2_z},\\
& a_i^6      = \sigma(z_i^6) \label{eq:cnn_fc2_a}. 
\end{alignat}
\end{subequations}

Equations (\ref{eq:cnn_conv1_z}) and (\ref{eq:cnn_conv1_a}) represent the first convolutional layer (Conv1 in Fig. \ref{fig:cnn_diagram}), while equation (\ref{eq:cnn_pool1}) represents the first max-pooling layer (Pool1 in Fig. \ref{fig:cnn_diagram}) with a $2\times2$ kernel. Equations (\ref{eq:cnn_conv2_z}) and (\ref{eq:cnn_conv2_a}) correspond to the second convolutional layer (Conv2 in Fig. \ref{fig:cnn_diagram}), and Equation (\ref{eq:cnn_pool1}) represents the second max-pooling layer (Pool2 in Fig. \ref{fig:cnn_diagram}) with a $2\times2$ kernel. Furthermore, equations (\ref{eq:cnn_fc1_z}) and (\ref{eq:cnn_fc1_a}) relate to the first fully connected layer (FC1 in Fig. \ref{fig:cnn_diagram}), and finally, equations (\ref{eq:cnn_fc2_z}) and (\ref{eq:cnn_fc2_a}) are associated with the second fully connected layer (FC2 in Fig. \ref{fig:cnn_diagram}). Here, Conv1 is referred to as layer 1, Pool1 as layer 2, Conv2 as layer 3, Pool2 as layer 4, FC1 as layer 5, and FC2 as layer 6. Following FC2, a softmax function normalizes the output into a probability distribution.

Now, the backpropagation functions for Fig. \ref{fig:cnn_diagram} are shown in equations (\ref{eq:cnn_bq_all}). In equation (\ref{eq:cnn_bp3}), the max-pooling layer is applied with case 1 represented by $a_{j,k}^4 = \text{max}(a_{2s,2t}^3, \ a_{2s+1,2t}^3, \ a_{2s,2t+1}^3,\ a_{2s+1,2t+1}^3)$, where $s=\frac{j}{2}$ and $t = \frac{k}{2}$. Similarly, in Equation (\ref{eq:cnn_bp4}), the $2\times2$ max-pooling layer is applied. 

Here, the softmax function is utilized as the activation function in the final layer, while the cross-entropy function is utilized as the loss function. This setting has found wide applications, such as image classification. There are two kinds of gradients: one in the fully connected layer, i.e., $\frac{\partial C}{\partial w_{i,k}^6}$, $\frac{\partial C}{\partial w_{i;j,k}^5}$, and another in the convolutional layer, $\frac{\partial C}{\partial w_{p,q}^3}$ and $\frac{\partial C}{\partial w_{p,q}^1}$. We will discuss each of them, respectively, in the following section. 

\begin{subequations} \label{eq:cnn_bq_all}
\begin{alignat}{2}
 \delta_i^6 &= \frac{\partial C}{\partial z_i^6} = \frac{\partial C}{\partial a_i^6}\frac{\partial a_i^6}{\partial z_i^6} = \frac{\partial C}{\partial a_6^3 }\sigma'(z_i^6) \label{eq:cnn_bp1},  \\
 \delta_i^5 &= \frac{\partial C}{\partial z_i^5} =  \sum_k^{N_6} \frac{\partial C}{\partial z_k^6} \frac{\partial z_k^6}{\partial a_i^5}\frac{\partial a_i^5}{\partial z_i^5}, \nonumber \\
 & = \sum_k^{N_6} \delta_k^6 w_{ki}^6 \sigma'(z_i^5), \label{eq:cnn_bp2} \\
 \delta_{j,k}^3 &= \frac{\partial C}{\partial z_{j,k}^3} = \sum_i \frac{\partial C}{\partial z_i^5} \frac{\partial z_i^5}{\partial a_{s,t}^4} \frac{\partial a_{s,t}^4}{\partial z_{j,k}^3}, \nonumber \\
 & = \sum_i \delta_i^5 w_{i;s,t}^5 \frac{\partial a_{s,t}^4}{\partial a_{j,k}^3} \frac{\partial a_{j,k}^3}{\partial z_{j,k}^3}, \nonumber \\
 & = \sum_i \delta_i^5 w_{i;s,t}^5 \frac{\partial a_{s,t}^4}{\partial a_{j,k}^3} \sigma'(z_{j,k}^3), \nonumber \\
 & = \begin{cases}
 \sum_i \delta_i^5 w_{i;s,t}^5 \sigma'(z_{j,k}^3), &\text{if case 1;} \\
0, &\text{otherwise;} 
\end{cases} \label{eq:cnn_bp3} \\
 \delta_{j,k}^1 &= \frac{\partial C}{\partial z_{j,k}^1} = \sum_u \sum_v \frac{\partial C}{\partial z_{u,v}^3} \frac{\partial z_{u,v}^3}{\partial a_{s,t}^2} \frac{\partial a_{s,t}^2}{\partial z_{j,k}^1} \nonumber , \\ 
 &= \sum_u \sum_v \delta_{u,v}^3 w_{u-s,v-t}^3 \frac{\partial a_{s,t}^2}{\partial a_{j,k}^1} \frac{\partial a_{j,k}^1}{\partial z_{j,k}^1},\nonumber \\
& = \sum_u \sum_v \delta_{u,v}^3 w_{u-s,v-t}^3 \frac{\partial a_{s,t}^2}{\partial a_{j,k}^1} \sigma'(z_{j,k}^1),  \nonumber \\
& = \begin{cases}
 \sum_u \sum_v \delta_{u,v}^3 w_{u-s,v-t}^3 \sigma'(z_{j,k}^1), &\text{if case 2;} \\
0, &\text{otherwise;} 
\end{cases} \label{eq:cnn_bp4} \\
 \frac{\partial C}{\partial w_{i,k}^6} & = \frac{\partial C}{\partial z_i^6} \frac{\partial z_i^6}{\partial w_i,k^6} = \delta_i^6 a_k^5, \label{eq:cnn_bp5}  \\
 \frac{\partial C}{\partial w_{i;j,k}^5} &= \frac{\partial C}{\partial z_{i}^5} \frac{\partial z_{i}^5}{\partial w_{i;j,k}^5} = \delta_i^5 a_{j,k}^4 \label{eq:cnn_bp6}, \\
 \frac{\partial C}{\partial w_{p,q}^3} &= \frac{\partial C}{\partial z_{j,k}^3} \frac{\partial z_{j,k}^3}{\partial w_{p,q}^3} = \delta_{j,k}^3 a_{j+p, \ k+q}^2 \label{eq:cnn_bp7}, \\
 \frac{\partial C}{\partial w_{p,q}^1} &= \frac{\partial C}{\partial z_{j,k}^1} \frac{\partial z_{j,k}^1}{\partial w_{p,q}^1} = \delta_{j,k}^1 x_{j+p, \ k+q}^0.\label{eq:cnn_bp8}
\end{alignat}
\end{subequations}

Calculating the gradient $\frac{\partial C}{\partial w_{i,k}^6}$ in the fully connected layer in equation (\ref{eq:cnn_bp5}) involves the multiplication of two terms $\delta_j^6$ and $a_k^5$. For $\frac{\partial C}{\partial w_{i;j,k}^5}$,
\begin{equation}
    \frac{\partial C}{\partial w_{i;j,k}^5} = [\sum_k^{N_6} \delta_k^6 w_{ki}^6 \sigma'(z_i^5)] a_{j,k}^4.
\end{equation}
The summation goes through all neurons in layer 6 and then by multiplication. 

For $\frac{\partial C}{\partial w_{p,q}^3}$ and $\frac{\partial C}{\partial w_{p,q}^1}$, we only consider nonzero conditions. 
\begin{equation} \label{eq:cnn_gradient_layer2}
\begin{aligned}
    \frac{\partial C}{\partial w_{p,q}^3}  & = \{ \sum_i \delta_i^5 w_{i;s,t}^5 \sigma'(z_{j,k}^3) \} a_{j+p, \ k+q}^2  \\
    & = \{\sum_i [\sum_k^{N_6} \delta_k^6 w_{ki}^6 \sigma'(z_i^5)] w_{i;s,t}^5 \sigma'(z_{j,k}^3) \} a_{j+p, \ k+q}^2
\end{aligned}
\end{equation}

\begin{equation}\label{eq:cnn_gradient_layer1}
\begin{aligned}
    \frac{\partial C}{\partial w_{p,q}^1}  & = \{ \sum_u \sum_v \delta_{u,v}^3 w_{u-s,v-t}^3 \sigma'(z_{j,k}^1) \} x_{j+p, \ k+q}^0  \\
    & = \{ \sum_u \sum_v [\sum_i \delta_i^5 w_{i;s,t}^5 \sigma'(z_{j,k}^3)] w_{u-s,v-t}^3 \sigma'(z_{j,k}^1) \}  \\ & \times x_{j+p, \ k+q}^0 \\
    & = \{ \sum_u \sum_v [\sum_i (\sum_k^{N_6} \delta_k^6 w_{ki}^6 \sigma'(z_i^5)) w_{i;s,t}^5  \sigma'(z_{j,k}^3)] \\ & w_{u-s,v-t}^3 \sigma'(z_{j,k}^1) \}  \times x_{j+p, \ k+q}^0
\end{aligned}
\end{equation} 

The gradient calculation in convolutional layers also involves multiplication and summation, while the gradient in the front layers involves more arithmetic operations.

\subsection{Gradient in Deep Neural Networks}
The summation and multiplication operations accumulate when neural networks deepen and the number of neurons in each layer increases. This can lead to gradient exploding problems \cite{exploding}, where the gradients grow exponentially as they are backpropagated through the neural network from layer to layer. As a consequence, vanilla SGD becomes unstable for model training. Several classical methods have been employed to address the issue of gradient exploding, including gradient clipping, proper weight initialization, and batch normalization. These techniques help control the magnitude of the gradients during the training process. Specifically, gradient clipping is a fast and effective method of limiting gradients to a predefined threshold. The threshold is introduced as an additional hyperparameter that needs to be selected carefully. When the threshold is set too low, the learning ability is restricted with small gradient updates. On the contrary, the clipping effect may be affected if the threshold is too high.

When the error $\delta_K^{l+1}$, the weight $w_{kj}^{l+1}$, and the derivative of the activation function $\sigma'(z_j^l)$ are small in equation (\ref{eq:bp2}), their summation can result in values in a much smaller range. Typically, the magnitude is much smaller than 1. This situation can lead to the gradient vanishing problem \cite{vanishing}, where the gradient exponentially diminishes as it is backpropagated to the earlier layers. Several techniques can be applied to mitigate the risk of gradient vanishing. Proper weight initialization and batch normalization are effective measures to reduce the probability of encountering this problem. However, replacing the sigmoid activation function with a Rectified Linear Unit (ReLU) is the most reliable approach. The ReLU function is defined as follows:
\begin{equation}
\sigma(x) = \left \{
  \begin{aligned}
    & 0, && x \leq 0; \\
    & x, && x > 0. 
  \end{aligned} \right. 
\end{equation}
\begin{equation}
\sigma'(x) = \left \{
  \begin{aligned}
    & 0, && x \leq 0; \\
    & 1, && x > 0. 
  \end{aligned} \right. 
\end{equation}
This results in the derivative of the activation function in (\ref{eq:bp2}) being 0 or 1. This adjustment enhances the likelihood of the activation function's derivative being 1, contributing to the stabilization of the neural network.

\section{Proposed Method}
Motivated by the gradient distribution and effective gradient clipping method to mitigate gradient exploding problems, we first present a received bit masking scheme to restrict the received gradient values within a small range. Then, approximate communication is applied by transmitting gradients with errors. The error level is measured using a $l_2$-norm. Lower transmission errors would result in better learning performance. Next, to enhance learning performance, the most significant bits (MSBs) in gradients are protected by gray coding. Finally, gradient compression is applied to further reduce communication costs in gradient transmission. 
\subsection{Received Bits Masking}
Section III provides mathematical demonstrations that establish that gradients can easily cause vanishing or exploding problems under specific conditions. Gradient clipping effectively mitigates the gradient exploding problem. This allows us to establish a threshold for the gradients received by the server. With this in mind, our approach starts by devising a reception mechanism tailored to these gradients.

\begin{figure}[ht]
	\includegraphics[width=3.2in]{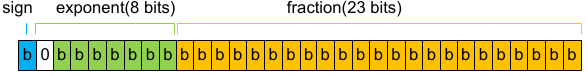}
	\centering
	\caption{Received Gradient Bit Representation}
	\label{fig:received_bit}
	\centering
\end{figure}

In ML, gradients are commonly represented by floating-point numbers of 32 bits. These numbers adhere to the structure dictated by the IEEE-754 standard. In this format, the initial bit is allocated for sign information, followed by 8 bits assigned to the exponent, and the concluding 23 bits designated for the fractional component. Each bit is susceptible to noise during transmission, resulting in potential corruption. To avoid block corruption, we integrate interleaving at the transmitter and deinterleaving. This strategy mitigates the probability of multiple error bits aligning together. Within the bit-level representation, if the second bit in the 32-bit form, i.e., the initial bit in the exponent section, is set to 1 while all other 31 bits are 0s, the corresponding decimal value becomes 2. Conversely, when the second bit is 0, and the other 31 bits are 1s, the value resides below 2. Assuming a magnitude threshold of 1 for the gradient value, the exponent's first bit is consistently set to 0. Building upon this insight, on the receiving end, irrespective of the decoded value in the second-bit position of the gradient, we mask it to 0 regardless of the actual received bit value. This adjustment is visually presented in Fig. \ref{fig:received_bit}, where the `0' bit signifies the constant value of 0 for that particular bit. In contrast, `b' adopts the value transmitted during reception.
\subsection{Approximate Wireless Transmission}
With the received bits masking described above, the gradient received at the server falls within the span of $(-2,2)$, as depicted in Fig. \ref{fig:received_bit}. Consequently, the associated error is also limited to a narrow interval. In practical scenarios, this range is often even more minor. In this context, the inherent error resilience of FL systems enables them to train models in the presence of errors. Specifically, gradients in FL systems can be transmitted with errors and need not be perfectly accurate. This resilience to error in FL systems manifests itself in two significant ways. First, the FL system operates within the domain of machine learning and is inherently equipped to manage gradient errors. Secondly, the global gradient aggregation process diminishes the magnitude of the error through averaging. To quantitatively evaluate the impact of errors, the researchers in \cite{quality_management} used relative error to assess individual value discrepancies. This relative error is precisely defined as:
\begin{equation}
    E_{rg} = \frac{|g - \Tilde{g} |}{g},
\end{equation}
where $g$ represents the original gradient value, $\Tilde{g}$ denotes the approximated value of $g$, and $E_{rg}$ stands for the relative error. Given the multitude of gradients present within the neural network, we employ the $l_2$-norm of errors instead of relative error to evaluate the impact of errors. The $l_2$-norm of errors is precisely characterized by the following:
\begin{equation} \label{eq:l2_error}
    || E_{r} ||_2 = \sqrt{\sum_{i=1}^{N_g} |g_i - \Tilde{g_i}|^2 },
\end{equation}
where $N_g$ is the number of gradients in the neural network.

We also adopt error tolerance as a metric to assess the threshold that yields comparable learning performance to accurate transmission. In this context, when $|| E_{r} ||_2 \leq E_T$, the learning performance remains minimally affected.

\subsection{Most Significant Bit Protection}
Furthermore, we have observed that the bits at different locations are of different significance and the gray coding within high-order modulations produces varying effects on bits located at different positions \cite{media}. This observation has significance not only in the transmission of media messages, but also in the transmission of ML models. In wireless communication, the transmission system lacks awareness of the relative importance of data bits and treats all bits equally. For example, when quadrature phase shift keying (QPSK) is used as the modulation scheme, each symbol comprises 2 bits, with possible combinations of ${00, 01, 11, 10}$. In QPSK, the error probability for the first and second bits is equivalent. In contrast, 16-QAM employs 4 bits per symbol, with a constellation map using gray coding, as illustrated below. As shown in Fig. \ref{fig:qam16_map}, one of the gray code mappings is depicted. Notably, it becomes evident that the smallest distance between two symbols corresponds to just a 1-bit difference.

\begin{figure}[ht]
	\includegraphics[width=1.8in]{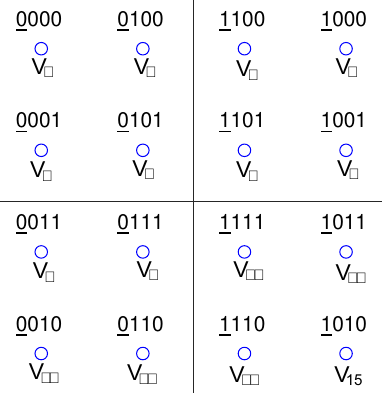}
	\centering
	\caption{Gray Coding in 16-QAM Constellation Map}
	\label{fig:qam16_map}
	\centering
\end{figure}

In Fig. \ref{fig:qam16_map}, the underlined bit corresponds to the first bit within each symbol, serving as the most significant bit (MSB) in the context of 16-QAM. Conversely, the fourth or last bit represents the least significant bit (LSB). When symbols share the same transmission probability, the error probability for the MSB is lower than that for the LSB. For example, if the symbol $s_0$ is decoded with an error, it will most likely be misconstrued as $s_1$, $s_4$, or $s_5$ when the noise power is minimal. For simplicity, we assume that the error symbol can only be ${s_1, s_4, s_5}$. While the MSB remains constant, the LSB changes twice. Assuming that the probability of symbol error from $s_0$ to $s_1$ is $\rho$, then the probability of symbol error for $s_4$ is $\rho$, and it becomes $\sqrt2\rho$ for $s_5$. This observation is summarized in Table II. The symbols within the other quadrants exhibit symmetry to that of the first quadrant, yielding identical results. In high-order modulations, gray coding safeguards the MSB bits of gradient values represented in bit form.

\begin{table}[h]
	\newcommand{\tabincell}[2]{\begin{tabular}{@{}#1@{}}#2\end{tabular}}
	\centering
	\caption{Gray Coding with 16-QAM MSB/LSB Error Count}
	\begin{tabular}{c|p{19mm}|p{23mm}|p{23mm}}
		\hline
		\textbf{Symbol} & \textbf{Potential Error Symbol} & \textbf{MSB Error Count/ Probability} & \textbf{LSB Error Count/ Probability} \\
		\hline
		$s_0$ & \tabincell{l}{$s_1, s_4, s_5$} & 0 / 0 & 2 / $(1+\sqrt{2})\rho$\\
		\hline
		$s_1$ & \tabincell{l}{$s_0, s_2, s_4, s_5, s_6$} & 2 / $(1+\sqrt{2})\rho$ & 3 / $(1+2\sqrt{2})\rho$\\
		\hline
		$s_4$ & \tabincell{l}{$s_0, s_1, s_5, s_8, s_9$} & 0 / 0 &  2 / $(1+\sqrt{2})\rho$ \\
		\hline
            $s_5$ & \tabincell{l}{$s_0, s_1, s_2, s_4$, \\ $s_6, s_8, s_9, s_{10}$} & 3 / $(1+2\sqrt{2})\rho$ & 3 / $(1+2\sqrt{2})\rho$\\
		\hline
	\end{tabular}
	\label{Tab:error_count}
\end{table}

\subsection{Gradient Compression}
Gradient compression can further reduce the communication cost. In \cite{noma}, we showed that gradient quantization and sparsification can greatly reduce communication time. This would cause a slight loss in final learning performance due to compression errors. In our proposed approximate transmission scheme, there exist transmission errors. We applied the sparsification here. Only the gradients with large gradient magnitude and their indices will be transmitted to the server. Gradient indices will be transmitted accurately through another channel. On the receiver side, the gradients at the specific positions will be set to zero. 

\subsection{Summary}
The whole process is described in Algorithm \ref{al:error_fl}. Our approach eliminates the need for FEC and packet retransmission. It is important to note that our methodology diverges from UDP, where retransmission is unnecessary. The distinction lies in the fact that UDP operates at a higher level, using the CRC solely to verify the UDP payload. Regarding physical or MAC layer errors, retransmission is still invoked within UDP. In contrast, our approach eliminates FEC and retransmission in the lower layers, encompassing the physical and MAC layers. This approach yields a tripartite benefit: First, it reduces communication overhead, facilitating the transmission of more data bits. Second, it mitigates the computational burden associated with FEC, a particularly attractive feature for edge devices. Third, it enhances latency performance, as retransmission becomes unnecessary.

\begin{algorithm}[]
\caption{Approximate Wireless Communication for Federated Learning} 
\begin{algorithmic}[1] \label{al:error_fl}
\STATE  {\bf Initialization:} {\bf Server} initializes $w_0$ and broadcasts to all {\bf Clients}.
\WHILE{not converge}
\STATE {\bf Client:}\\
    \quad Receives the most recent aggregated global model. \\
    \quad Performs local computation as Eq. (\ref{eq:local_comp}). \\
    \quad Sends local gradient directly without error correction coding through wireless channels.
\STATE {\bf Server:} \\
    \quad Receives the local gradient with errors.\\
    \quad Masks the second bit in the 32-bit gradient representation as 0, leaving the other bits as received. \\
    \quad Sends the aggregated global model to clients.
\ENDWHILE
\end{algorithmic}
\end{algorithm}

\section{Simulation Results}
This section begins by presenting the configurations of the simulation parameters. After that, we conducted BER simulation under different modulation schemes. Notably, among the modulation schemes, the lower-order modulation technique exhibits a superior BER to higher-order modulation, given the same SNR level. Specifically, QPSK outperforms 16-QAM and 256-QAM in terms of BER. Moving forward, we can compare FL performance across three scenarios: ECRT, naive erroneous transmission, and erroneous message augmented by our proposed strategy. The naive erroneous transmission signifies that transmission in wireless networks is subject to errors without supplementary measures. In contrast, our proposed scheme shows significantly improved testing accuracy compared to naive erroneous transmission. Compared to ECRT transmission, our proposed method yields substantial time savings while achieving commendable performance. The convergence of our proposed scheme is proved using three different image datasets in both IID and non-IID data settings. Then, the aggregation error is quantized and the error is evaluated under different wireless conditions. Furthermore, we dive into the implications of gray coding across various modulation schemes, highlighting the innate bit protection for MSBs. Finally, we investigate the influence of user participation numbers on the FL process. As user participation increases, aggregation errors decrease, indicating a pronounced improvement. 

Our simulation considers a prototypical FL setup with $M=100$ clients that interact with the server. We leverage three image datasets to validate our proposed approach: MNIST, Fashion-MNIST, and Cifar-10. These datasets encompass $10$ distinct classes, and while MNIST and Fashion-MNIST entail grayscale images, Cifar-10 features color images. Specifically, MNIST comprises handwritten digits from 0 to 9, Fashion-MNIST showcases fashion articles like shoes and clothing, and Cifar-10 encompasses images of vehicles and animals. MNIST and Fashion-MNIST provide 60,000 images for training, whereas Cifar-10 offers 50,000. The test subsets for all datasets contain 10,000 images. Our simulation encapsulates Independently and Identically Distributed (IID) and non-IID data scenarios. In the IID setup, each client randomly samples a fixed number of images from the training data. Conversely, in the non-IID setup, each client is allocated images from 2 specific classes. Across the three datasets, CNNs are employed with varying architectures. The CNN configuration is similar to the datasets, incorporating $2$ convolutional layers with 5-unit kernels, $2$ max-pooling layers with a size of 2, and multiple fully connected layers in conclusion. The activation function in all layers, except the final one, uses ReLU, while the last layer uses the log-softmax function. In particular, the neural network architectures in the MNIST, Fashion-MNIST, and Cifar-10 datasets encompass 21,840, 65,558, and 62,006 model parameters, respectively. The learning rate is $\eta=0.01$, and the size of the training batch is 10. Throughout the learning process, testing follows each round after global aggregation, covering all data samples in the test data set. All clients participate in the training procedure, unless otherwise specified.

In the communication model, we establish the path loss exponent for the wireless channel as $\alpha=3$ and assume a distance of $10$ meters between the server and the clients. The transmission power at the clients is normalized to 1. 

QPSK achieves the best BER performance. The BER is summarized in Table \ref{Tab:ber}.
\begin{table}[h]
	\newcommand{\tabincell}[2]{\begin{tabular}{@{}#1@{}}#2\end{tabular}}
	\centering
	\caption{BER Summary}
	\begin{tabular}{c|p{19mm}|p{19mm}|p{19mm}}
		\hline
		\textbf{SNR} & \textbf{QPSK} & \textbf{16-QAM} & \textbf{256-QAM} \\
            \hline
		0 dB & $2.11\times10^{-1}$ & $3.28\times10^{-1}$ & $4.26\times10^{-1}$\\
		\hline
		10 dB & $4.36\times10^{-2}$ & $1.23\times10^{-1}$ & $2.79\times10^{-1}$\\
		\hline
		20 dB & $4.91\times10^{-3}$ & $1.90\times10^{-2}$ & $1.12\times10^{-1}$\\
		\hline
	\end{tabular}
	\label{Tab:ber}
\end{table}

\begin{figure}[ht]
  \centering
  \subfloat[Test Accuracy of IID Data at SNR=10 dB]{\includegraphics[width=0.24\textwidth]{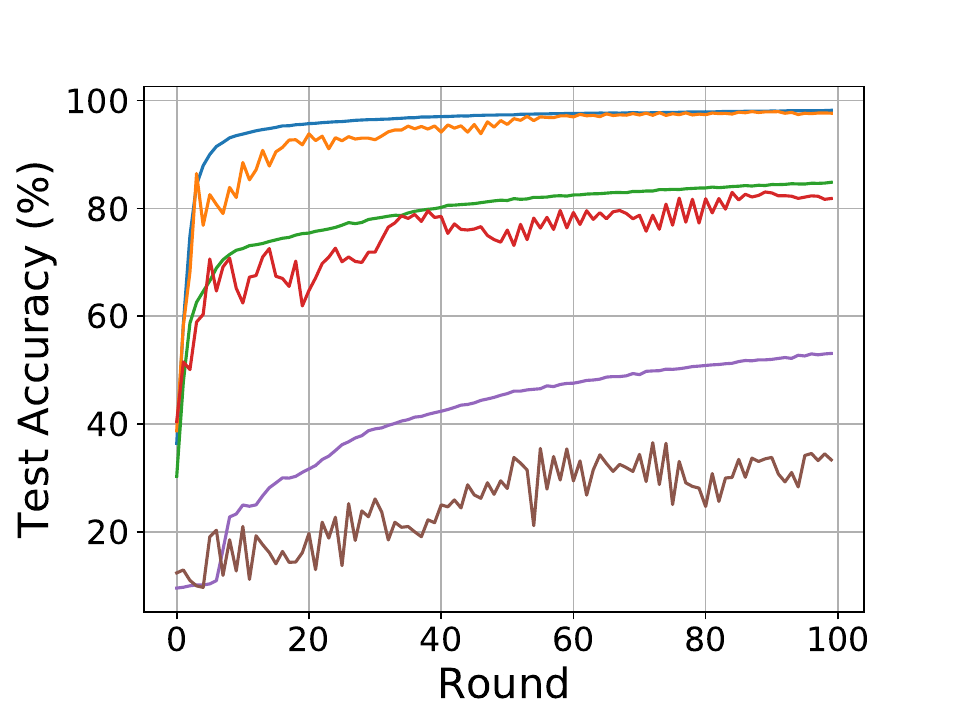}\label{fig:f1}}
  \hfill
  \subfloat[Test Accuracy of Non-IID Data at SNR=20 dB]{\includegraphics[width=0.24\textwidth]{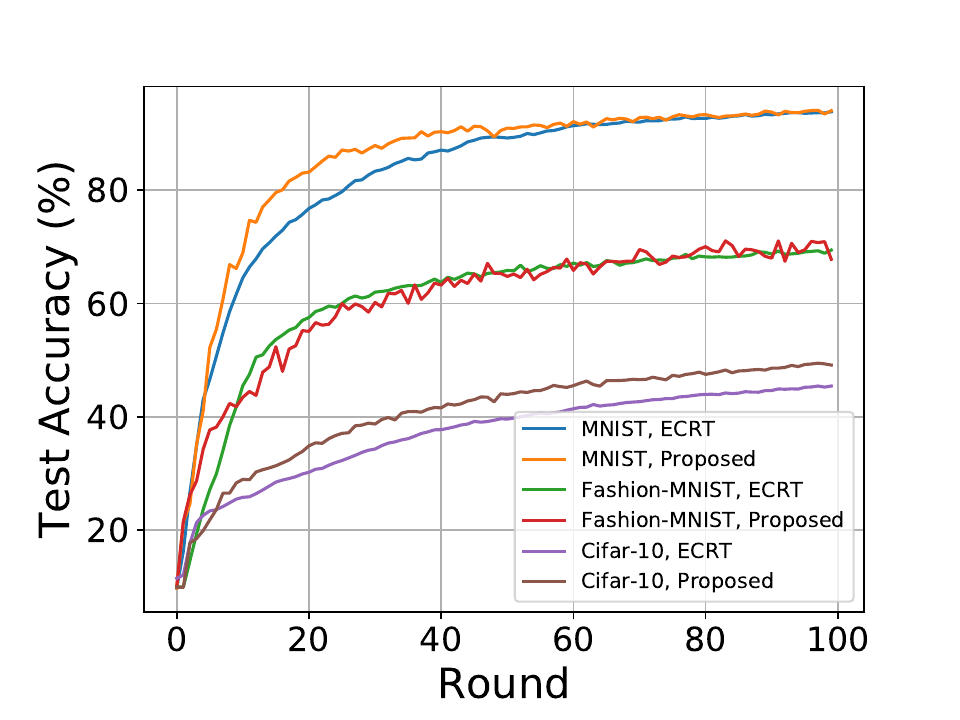}\label{fig:f2}}
  \caption{Test Accuracy}
  \label{fig:convergence}
\end{figure}

To begin with, we demonstrate that the proposed scheme attains a convergence point similar to that of the ECRT scheme under the high SNR regime while experiencing a marginal reduction in performance under the low SNR regime. Fig. \ref{fig:convergence} illustrates the learning outcomes for three different data sets, comparing the transmission of ECRT with our proposed transmission method using QPSK modulation. We use the IID data set in Fig. \ref{fig:convergence}(\subref{fig:f1}). When our proposed method is used, the learning curve exhibits a more significant fluctuation than that observed with ECRT transmission. This variance is attributed to transmission errors introduced by approximate communication. However, except for the Cifar-10 dataset, the convergence point of learning of our proposed method aligns closely with that of ECRT transmission for the other two datasets. This fluctuation becomes more streamlined in the high SNR regime shown in Fig. \ref{fig:convergence}(\subref{fig:f2}). Furthermore, due to the non-IID data structure, our proposed method showcases improved learning performance in certain instances, surpassing that of ECRT transmission. 

\begin{figure}[ht]
	\includegraphics[width=0.4\textwidth]{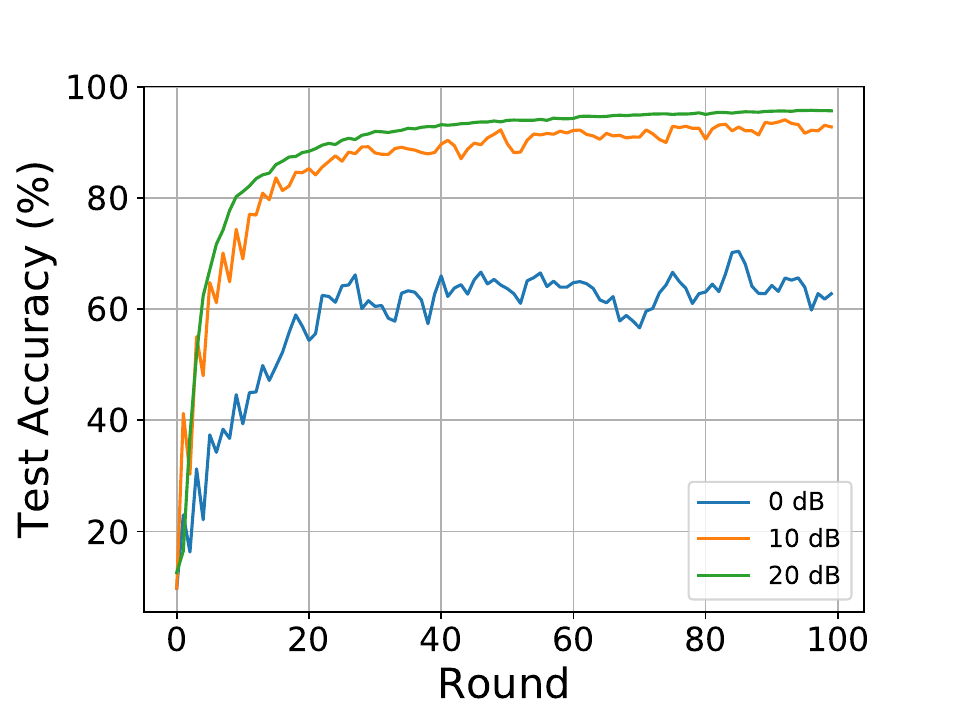}
	\centering
	\caption{Test Accuracy of MNIST, Non-IID with Different SNR}
	\label{fig:convergency_channel}
	\centering
\end{figure}

Wi-Fi transmission nominally targets the SNR range of 10-30 dB \cite{wifi_snr}. When SNR=0 dB, there will be no valid packet due to transmission errors and packet retransmission. Our proposed method still achieves about $60\%$ test precision when SNR = 0 dB, as shown in Fig. \ref{fig:convergency_channel}. The proposed method is superior to the ECRT method with a very low SNR. In addition, a high SNR gets better test accuracy than a low SNR.

\begin{figure}[ht]
  \centering
  \subfloat[$l_2$ error norm at SNR =10 dB]{\includegraphics[width=0.24\textwidth]{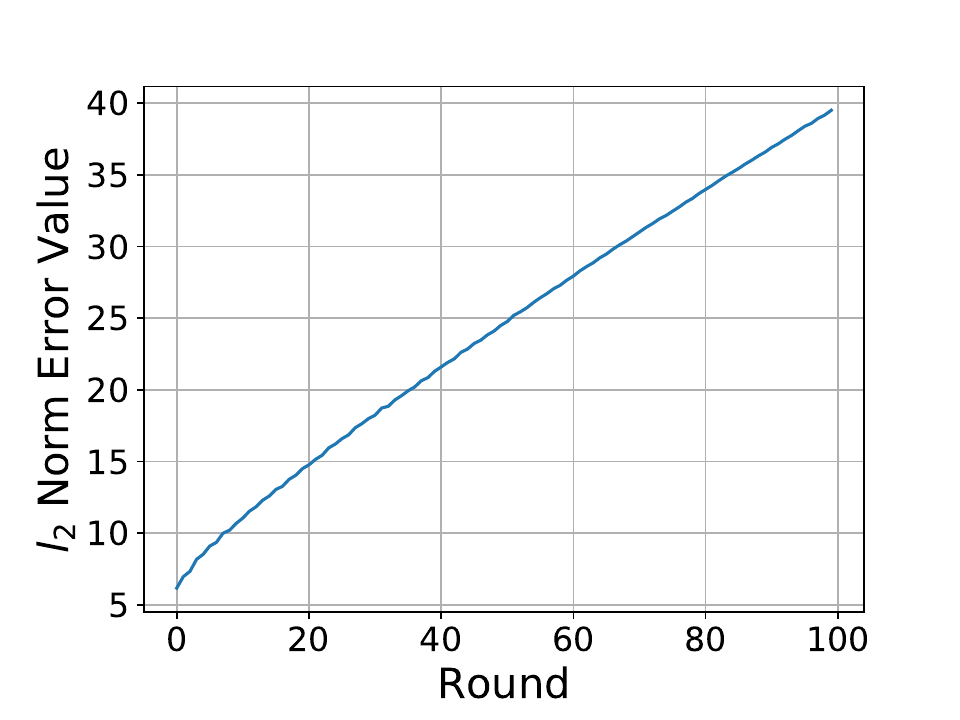}\label{fig:single_error_norm}}
  \hfill
  \subfloat[Average $l_2$ error norm at different SNR]{\includegraphics[width=0.24\textwidth]{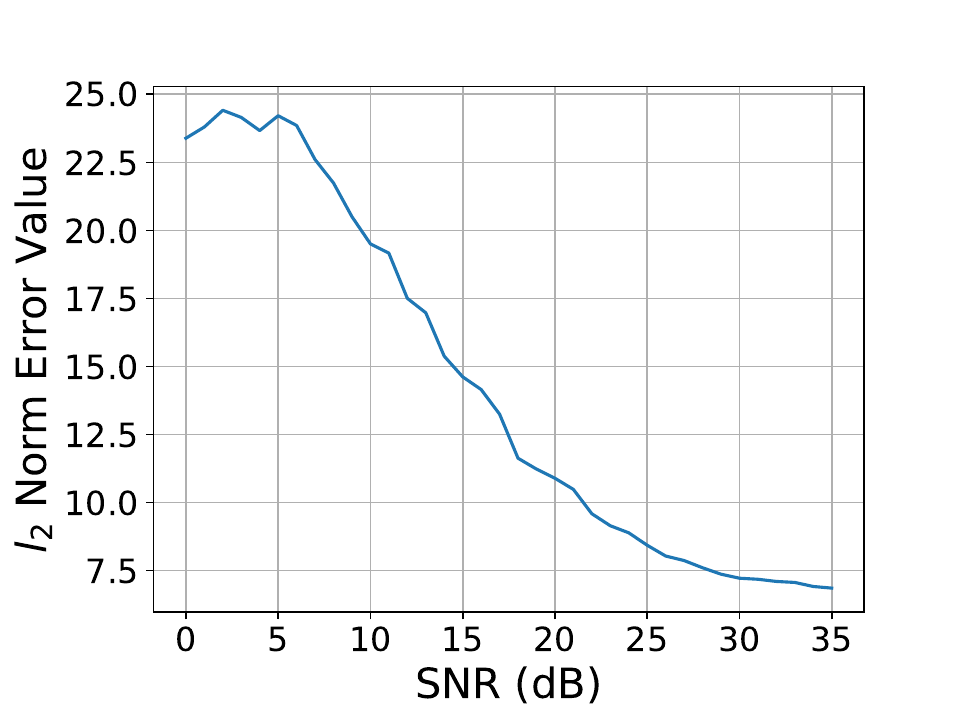}\label{fig:avg_error_norm}}
  \caption{$l_2$ error norm}
  \label{fig:error_level}
\end{figure}

To quantify the impact of errors, we employ the $l_2$-norm of the gradient aggregation error as our evaluation metric, as depicted in Equation (\ref{eq:l2_error}). Fig. \ref{fig:error_level}(\subref{fig:single_error_norm}) shows the error values of the norm gradient $l_2$ when using our proposed method with 256-QAM in the MNIST dataset under non-IID conditions at SNR=10 dB. In particular, the error exhibits a linear accumulation pattern. To assess the influence of SNR on the aggregation error, we employ average aggregation as our measure. For each level of SNR, we run 100 rounds and calculate the average value of the $l_2$ norm error. Fig. \ref{fig:error_level}(\subref{fig:avg_error_norm}) illustrates that the average error decreases with higher SNR values. This observation explains why the learning performance is superior at SNR=20 dB compared to SNR=10 dB, as demonstrated in Fig. \ref{fig:convergence}. 

\begin{figure}[ht]
	\includegraphics[width=0.4\textwidth]{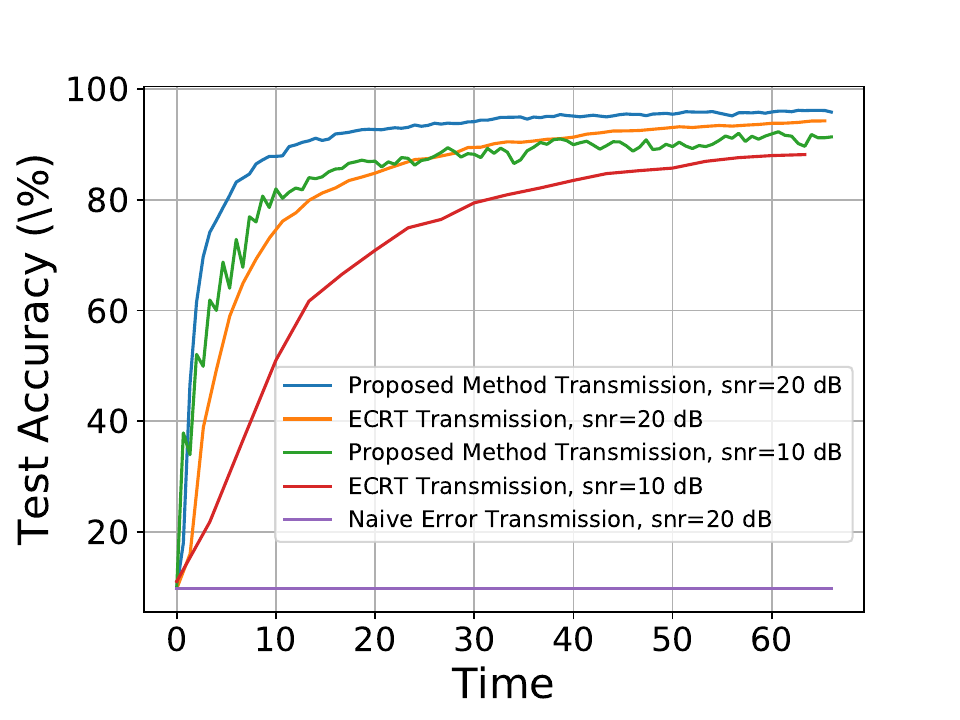}
	\centering
	\caption{Test Accuracy of MNIST under Non-IID}
	\label{fig:mnist_niid_time}
	\centering
\end{figure}

Within the ECRT scheme, reliable bit reception at the server is ensured through error correction coding and retransmission mechanisms, though at the expense of time and energy. Conversely, the naive and erroneous transmission approach involves transmitting flawed bits without gaining insight into gradient values. This leads to a test accuracy that remains flat at approximately $10\%$, akin to random guessing, as depicted in Fig. \ref{fig:mnist_niid_time}. On the other hand, our proposed method leverages prior knowledge of gradient values, which are anticipated to be distributed in a small range. As a result, our proposed scheme outperforms naive error transmission, resulting in significantly improved outcomes.

To quantify the time savings achieved by our proposed method compared to ECRT transmission, we employ a practical IEEE 802.11 protocol coupled with LDPC error correction coding. LDPC is a promising error correction code that approaches the Shannon limit. The choice of coding rate introduces a trade-off between error correction capability and transmission overhead. Lower coding rates increase transmission overhead, but offer enhanced error correction capabilities. Our approach employs a coding rate of $1/2$, which is used to improve error correction. As reported in \cite{ldpc}, for a code rate of $1/2$ with a code length of 648 and the use of the parity check matrix for the search, the minimum Hamming distance amounts to 15. This translates to an error correction capacity of 7 bits. ECRT transmission using QPSK modulation entails retransmission rates of $3.4\%$ and $23\%$ at SNR=20 dB and SNR=10 dB, respectively. Retransmission of the packet will reduce TCP throughput \cite{tcp_throughput}. As demonstrated in Fig. \ref{fig:mnist_niid_time}, the use of LDPC coding and retransmission extends the time required to achieve test precision $80\%$ to twice that of the proposed scheme at SNR=20 dB. Consequently, this difference increases to more than threefold at SNR = 10 dB, highlighting the advantage of the proposed method. 

\begin{figure}[ht]
  \centering
  \subfloat[Test Accuracy at the Same SNR=10 dB]{\includegraphics[width=0.24\textwidth]{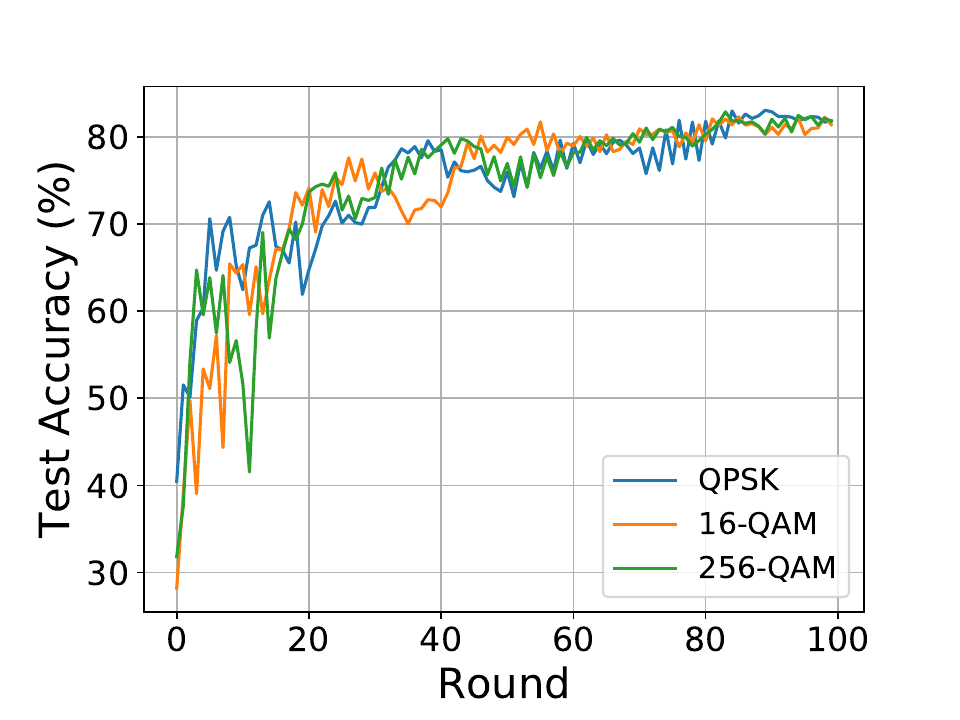}\label{fig:snr}}
  \hfill
  \subfloat[Test Accuracy at the Same BER $\approx 4.36 \times 10^{-2}$]{\includegraphics[width=0.24\textwidth]{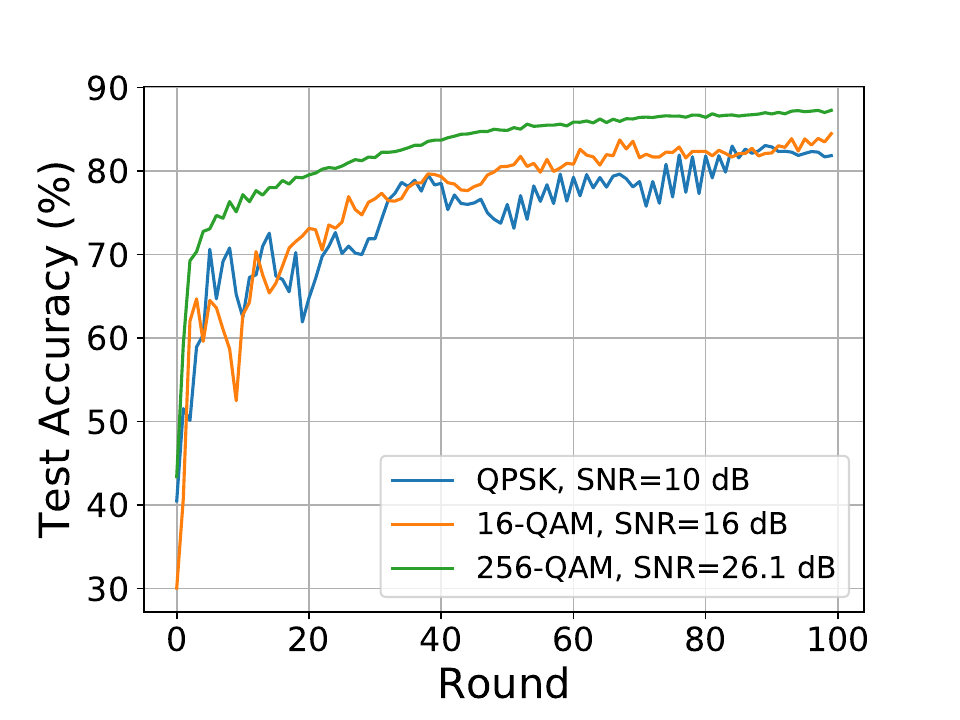}\label{fig:ber}}
  \caption{Test Accuracy of Fashion-MNIST under IID}
  \label{fig:same_snr_ber}
\end{figure}

To show the effectiveness of inherent MSB bit protection offered by gray coding in high-order modulation, we start by presenting the test accuracy of distinct modulations at an equivalent SNR in Fig. \ref{fig:same_snr_ber}(\subref{fig:snr}). Fashion-MNIST attains comparable learning performance when modulated using QPSK, 16-QAM, and 256-QAM, all at an SNR of 10 dB. This underlines the resilience conferred by gray coding in varying modulation schemes.

\begin{table}[h]
	\newcommand{\tabincell}[2]{\begin{tabular}{@{}#1@{}}#2\end{tabular}}
	\centering
	\caption{SNR for Target BER}
	\begin{tabular}{c|p{19mm}|p{19mm}|p{19mm}}
		\hline
		\textbf{BER} & \textbf{QPSK} & \textbf{16-QAM} & \textbf{256-QAM} \\
		\hline
		$4.36\times10^{-2}$ & 10 dB & 16 dB & 26.1 dB\\
		\hline
		$4.91\times10^{-3}$ & 20 dB & 26.1 dB & 36.5 dB\\
		\hline
	\end{tabular}
	\label{Tab:snr}
\end{table}

Table \ref{Tab:ber} presents the target SNR values required for different modulations to obtain an identical BER. Although QPSK, 16-QAM, and 256-QAM achieve similar BER levels as depicted in Fig. \ref{fig:same_snr_ber}(\subref{fig:ber}), it is noteworthy that the learning performance of gray-coded 256-QAM transmission surpasses that of both QPSK and 16-QAM. This observation underscores the enhanced performance obtained by employing gray coding in conjunction with 256-QAM modulation.

\begin{figure}[ht]
  \centering
  \subfloat[Test Accuracy Using ECRT Transmission]{\includegraphics[width=0.24\textwidth]{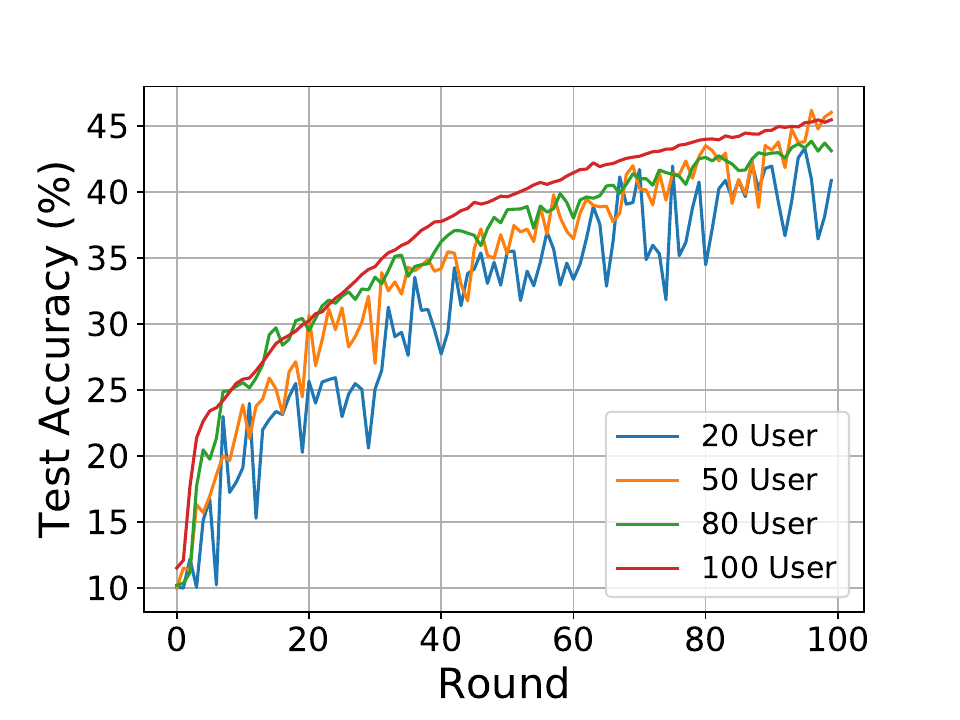}\label{fig:cifar_user_no_channel}}
  \hfill
  \subfloat[Test Accuracy Using Our Proposed Method]{\includegraphics[width=0.24\textwidth]{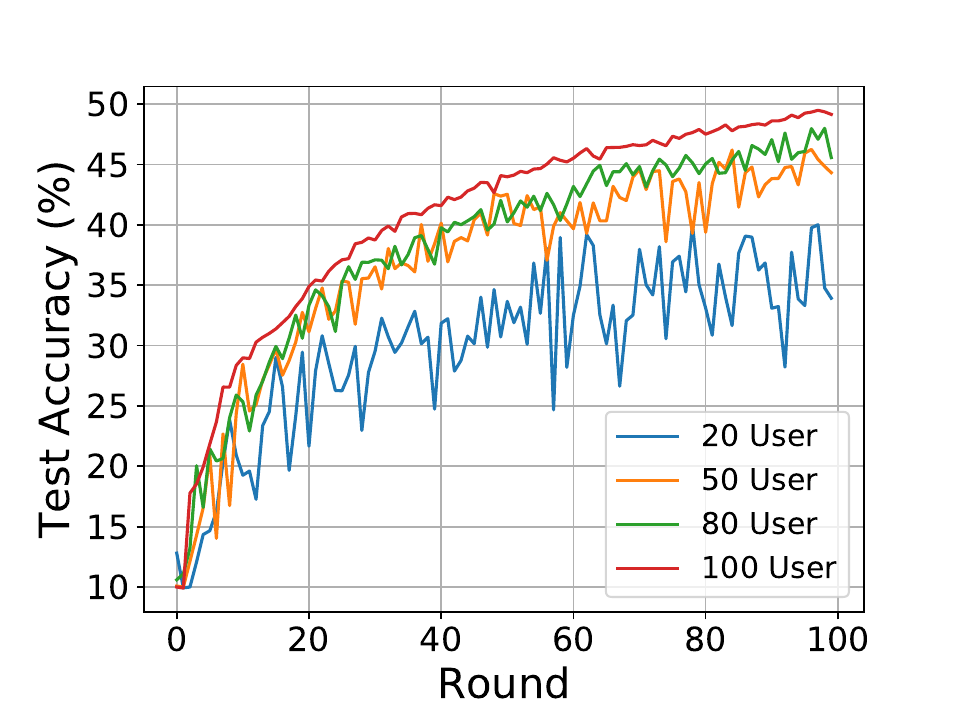}\label{fig:cifar_user_m4_20db}}
  \caption{Test Accuracy of Cifar-10 under non-IID}
  \label{fig:cifar_user}
\end{figure}
In the context of FL, model averaging is employed for aggregation, which inherently mitigates error levels. As demonstrated in Fig. \ref{fig:cifar_user}(\subref{fig:cifar_user_no_channel}), the participation of 20, 50, 80, or 100 users in the learning process, free of errors, produces comparable learning outcomes. In contrast, using QPSK modulation at SNR=20 dB in Fig. \ref{fig:cifar_user}(\subref{fig:cifar_user_m4_20db}), when 20 users engage in learning with errors introduces more pronounced fluctuations. However, when the participant count increases to 50, 80, or 100, the behavior mirrors that of the ECRT transmission. By involving more users in our proposed method, the expected outcome is reduced error and improved performance. 

\begin{figure}[ht]
	\includegraphics[width=0.48\textwidth]{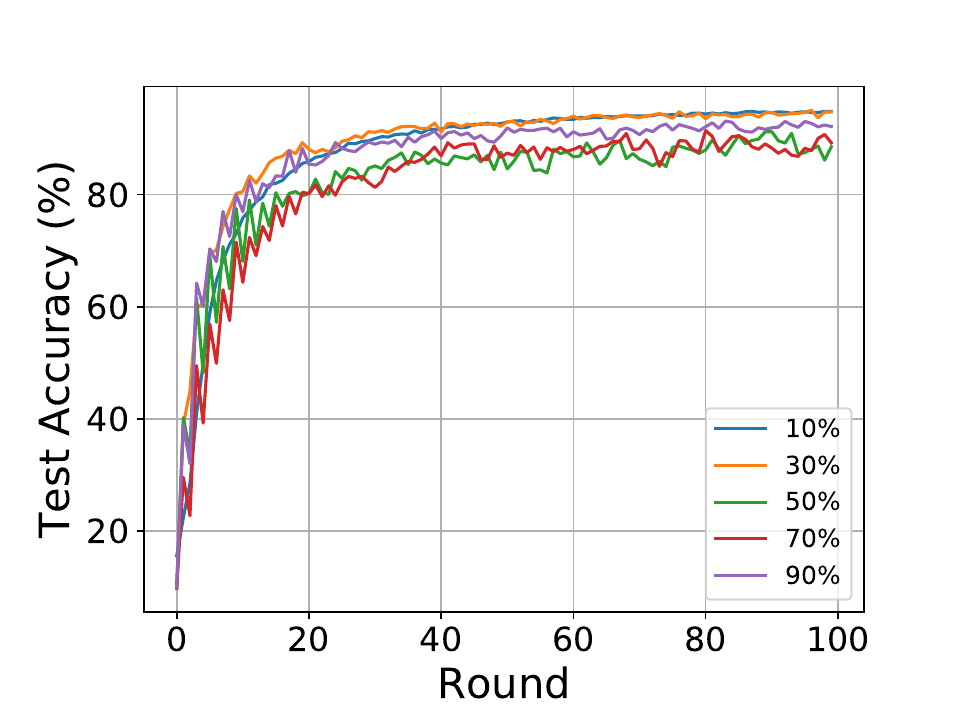}
	\centering
	\caption{Test Accuracy of MNIST under Non-IID with Approximate Communication and Sparsification}
	\label{fig:sparsification_round_niid}
	\centering
\end{figure}

The effectiveness of gradient compression is shown from the perspective of the number of required rounds. When the gradients are sparsified to $10\%$, $30\%$, $50\%$, $70\%$, $90\%$, respectively, the final learning performance is shown in Fig. \ref{fig:sparsification_round_niid}. From the results, when only $10\%$ of the gradients are kept, the best learning performance is achieved. In approximate communication, the gradients with errors can also be destructive rather than constructive. From a time perspective, when the sparsification rate is $10\%$, while the indices of the kept gradients need to be transmitted in accurate channels, compression still saves close to $10\times$ time.

\section{Conclusion}
This paper introduces a novel approach to transmitting model parameters in federated learning within wireless networks for resource-constrained IoT devices. Unlike existing transmission methods that rely on forward error correction and retransmission, our proposed scheme involves gradient transmission with errors leveraging prior knowledge of the gradient values. This knowledge is based on the gradient distribution. By masking the bits received, our approach achieves significantly improved learning performance with errors, outperforming naive error transmission. In particular, it achieves learning performance comparable to conventional transmission with forward error correction and packet retransmission, while consuming at least half the time. Furthermore, this study investigates the impact of factors such as the variance of the aggregation error with SNR, the effectiveness of gray coding in high-order modulation, and the influence of the number of users participating in the learning process under error conditions. Gradient sparsification can further reduce communication costs and mitigate the side effects caused by approximate gradient transmission.

\vfill

\end{document}